\documentclass[10pt, a4paper]{article}
\usepackage{amsmath,amssymb,amsthm,times,graphics,graphicx,bm}

\usepackage[colorlinks,linkcolor=red,citecolor=blue,urlcolor=blue]{hyperref}

\newcommand{\be}{\begin{equation}}
\newcommand{\ee}{\end{equation}}
\newcommand{\ben}{\begin{eqnarray}}
\newcommand{\een}{\end{eqnarray}}
\newcommand{\bes}{\begin{subequations}}
\newcommand{\ees}{\end{subequations}}
\newcommand{\bF}{\begin{figure}}
\newcommand{\eF}{\end{figure}}

\def\ket#1{ | #1 \rangle}

\def\ev#1{\langle #1 \rangle}

\def\pd2v#1#2#3{\frac{\partial^2 #1}{\partial #2 \partial #3}}

\def \2x2mat#1#2#3#4{
\left( \begin{array}{cc}
#1 &  #2 \\  #3 &  #4
\end{array} \right)
}

\title{Anomalous Weak Values and the Violation of a Multiple-measurement Leggett-Garg Inequality}

\author{Alessio Avella\\
{\small   Istituto Nazionale di Ricerca Metrologica, Strada delle Cacce 91, 10135, Torino, Italy}\\
\and 
Fabrizio Piacentini\\
{\small   Istituto Nazionale di Ricerca Metrologica, Strada delle Cacce 91, 10135, Torino, Italy} \\
\and 
Michelangelo Borsarelli \\
{\small Universit\`a degli Studi di Torino, via P. Giuria 1, 10125, Torino, Italy} \\ \and 
Marco Barbieri \\
{\small Universit\`a degli Studi Roma Tre, Via della Vasca Navale 84, 00146 Rome, Italy} \\ \and 
Marco Gramegna \\
{\small   Istituto Nazionale di Ricerca Metrologica, Strada delle Cacce 91, 10135, Torino, Italy} \\\and 
Rudi Lussana\\
{\small Politecnico di Milano, Piazza Leonardo da Vinci 32, 20133 Milano, Italy}\\ \and 
Federica Villa\\
{\small Politecnico di Milano, Piazza Leonardo da Vinci 32, 20133 Milano, Italy}\\ \and 
Alberto Tosi\\
{\small Politecnico di Milano, Piazza Leonardo da Vinci 32, 20133 Milano, Italy}\\ \and 
Ivo Pietro Degiovanni \\
{\small   Istituto Nazionale di Ricerca Metrologica, Strada delle Cacce 91, 10135, Torino, Italy} \\ \and 
Marco Genovese \\
{\small   Istituto Nazionale di Ricerca Metrologica, Strada delle Cacce 91, 10135, Torino, Italy} \\
	}

\begin{document}

\maketitle

\begin{abstract}
Quantum mechanics presents peculiar properties that, on the one hand, have been the subject of several theoretical and experimental studies about its very foundations and, on the other hand, provide tools for developing new technologies, the so-called quantum technologies.
The nonclassicality pointed out by Leggett-Garg inequalities has represented, with Bell inequalities, one of the most investigated  subject.
In this letter we study the connection of Leggett-Garg inequalities with a new emerging field of quantum measurement, the weak values.
In particular, we perform an experimental study of the four-time correlators Legget-Garg test, by exploiting single and sequential weak measurements performed on heralded single photons.
We show violation of a four-parameters Leggett-Garg inequality in different experimental conditions, demonstrating an interesting connection between Leggett-Garg inequality violation and anomalous weak values.
\end{abstract}

{\it Introduction.} The prominent role of measurement is one of the distinctive features of quantum theory \cite{mg1}. The impossibility of interpreting the results of a measurement on a quantum system in terms of pre-existing values is the core message of Bell's nonlocality test~\cite{Bell64,mg}, as well as the one of non-contextuality tests~\cite{KS39}. Such occurrence has also been recognised by Leggett and Garg in the behaviour of macroscopic systems when subject to subsequent measurements~\cite{LG85}.
For these objects, it is natural to assume that they will be found in a definite, realistic macro-state ({\it macroscopic realism}), and that a measurement, especially when carried out by a microscopic probe, can not perturb such macro-state ({\it non-invasive measurability}).
This original observation by Leggett and Garg has lead to a fecund production of theoretical~\cite{bruckner, bruckner2, mm,Lambert, Nimmrichter, BuEmary,Robens,Morgan} and experimental~\cite{Palacios, Goggin, Dressel, Athalye, Knee, Holger,Huelga, Knee2} work focussing on the inadequacy of such macro-realistic view~\cite{Emary14}; this has also inspired somehow the transposition of Bell's nonlocal argument to the time domain~\cite{Vedral, Fritz,Fedrizzi,Budroni,Brierley}.

In its simplest form, Leggett and Garg's arrangement considers a macroscopic body undergoing three two-outcome measurements at different times, with the first serving as a preparation. The correlation among the outcomes can be shown not to be in accordance with macro-realistic prescriptions. To date, the violation of the Leggett-Garg inequality has been reported on macroscopic objects, such as transmon qubits~\cite{Palacios} and crystals~\cite{Huelga}, and following refinements have been explored with superconducting flux qubits~\cite{Knee2}. The test, however, is also suited to highlight the inadequacy of a realistic view to the description of simpler quantum objects, such as phosphorus impurities~\cite{Knee}. In this case, the focus is rather in the assessment of the quantum character of the system in view of technical applications, than in its fundamental value.

The canonic three-measurement arrangement can be generalised in several directions. The simplest extension considers longer sequences~\cite{bruckner, mm}, and can lead to larger discrepancies, as it has been tested with photons~\cite{china} and nuclear spins~\cite{Athalye}. A different take considers substituting the measurement in the middle with a weak measurement imparting limited back-action~\cite{weak1}: while shot by shot the measurement delivers only partial information on the observable, it still provides the correct expectation value on a large ensemble~\cite{Wiseman}. This concept has been introduced in~\cite{Jordan1, Jordan}, and tested on single photons in~\cite{Goggin}, with further extension to multi-party scenarios appearing shortly after~\cite{Dressel}.

The experimental scheme for a Leggett-Garg test (LGT) can also be employed for observing so-called post-selected values: the value of the second observable is considered only on events chosen according to the outcome of the last measurement.  Post-selection procedures are expected to be mostly harmless in classical statistics, although the subject is vigorously debated~\cite{Ferrie,Brodutch,FerrieR,Aharonov}; in this context, post-selection operated in the weak-measurement regime can lead to anomalous values, in that they fall outside the range allowed to standard values~\cite{AharonovS,Pryde}. When one allows for such a weak measurement to be performed in a Leggett-Garg test, then a direct connection can be established between the violation of macro-realism and the emergence of anomalous post-selected values~\cite{Jordan1,Goggin,JordanErr}, like it was demonstrated for quantum contextuality \cite{pusey,pusey2}.

In this Letter, we present an experiment encompassing these two generalisations at the same time, by demonstrating a multiple-measurement setting operated in the weak regime. We perform a LGT on the polarisation of single photons, estimating non-commuting observables via 'weak averages'~\cite{Piacentini}, and draw an explicit link to the emergence of anomalous values. Our experiment confirms the intimate connection between the observation of anomalies in the post-selected statistics of quantum measurement, and the failure of a macrorealistic view.\\

{\it Theoretical background.} The simplest LGT one can design involves three measurements, which we label as $I_A, I_B$, and $I_C$; these are two-outcome observables which can take either the value +1 or -1. The inequality writes~\cite{LG85}:
\begin{equation}
-3\leq \mathcal{B}_3=\ev{I_A I_B}+\ev{I_B I_C}-\ev{I_A I_C}\leq1.
\end{equation}
The measurement of $I_A$ can be taken to coincide with the initial preparation in the state $\ket{\psi_A}$~\cite{Jordan}, hence one can assign the fixed value +1 for $I_A$: \begin{equation}
-3\leq\mathcal{B}_3=\ev{I_B}+\ev{I_B I_C}-\ev{I_C}\leq1.
\end{equation} The connection with anomalous post-selected values of $I_B$ is established by considering the two instances $I_C=1$ and $I_C=-1$ separately, each with the respective occurrence probabilities $p_C(1)$ and $p_C(-1)$:
\begin{equation}
\begin{aligned}
\label{LG3}
&\mathcal{B}_3=\ev{I_B}+\big[{_1\ev{I_B}}-1\big]p_C(1) - \big[{_{-1}\ev{I_B}}-1\big]p_C(-1)
\end{aligned}
\end{equation}
with ${_a\ev{I_B}}$ identifying the post-selected value of $I_B$, conditioned on the outcome $a$ for $I_C$.
Exploiting the relation:\begin{equation}
\begin{aligned}\ev{I_B} = {_1\ev{I_B}} p_C(1) + {_{-1}\ev{I_B} p_C(-1)}, \end{aligned}
\end{equation}
it is possible manipulate the Eq. \eqref{LG3} as
\begin{equation}
\begin{aligned}
\label{limit3}
&\mathcal{B}_3=1+2p_C(1)\left({_1\ev{I_B}}-1\right)
\end{aligned}
\end{equation}
 Inserting the condition for the standard values of $\ev{I_B}$, one recovers the limits of the Leggett-Garg inequality.

This connection can be extended to the multiple-measurement LGT introduced in~\cite{bruckner} that considers four measurements, including state preparation $I_A$:
\begin{equation}
\begin{aligned}
\label{LG4}
|\mathcal{B}_4|=\left|\ev{I_B}+\ev{I_B I_C}+\ev{I_C I_D}-\ev{I_D}\right|\leq2
\end{aligned}
\end{equation}
The form of this inequality resembles the familiar  Clauser-Holt-Shimony-Horne test for space-like separated systems~\cite{CHSH}; in that case, two partners alternate four distinct experimental arrangements, and verify whether the collected statistics is compatible with a local, realistic theory~\cite{Vedral,Fedrizzi}.
This can be viewed as a single system interrogated at four different times, including preparation. We can manipulate the four-measurement term $\mathcal{B}_4$ as we did for its three-measurement counterpart, by distinguishing the two instances for the last measurement $I_D$:
\begin{equation}
\begin{aligned}
\label{limit4}
|\mathcal{B}_4|=\left| \ev{I_B}+{\ev{I_BI_C}} + p_D(1)\left[{_1\ev{I_C}}-1\right]-\right.\\
-\left.p_D(-1)\left[{_{-1}\ev{I_C}}-1\right]\right|
\end{aligned}
\end{equation}
We now assume that the post-selected values are bound to be found in the same ranges as the standard values: in this case, it is easy to verify that $|\mathcal{B}_4|$ is upper bounded by 2. Differently from the three-measurement case, where any anomalous value would result in a violation, it can be shown that the inequality \eqref{limit4} demands a minimal value ${_-{\langle I_C \rangle}}\geq\frac{3-M}{2p_D(-1)}$, where $M=\ev{I_B}+{\ev{I_BI_C}}+\ev{I_C}$, with a similar expression holding for ${_+{\langle I_C \rangle}}$.\\

{\it Experimental implementation.} We perform a test of the inequality (\ref{limit4}) by exploiting single photons undergoing single and sequential weak measurements of their polarization. Single photons are emitted by a downconversion source~\cite{IvoOE}; at a heralding rate around $130$ kHz, the quality of the emission is certified by a measured value of the antibunching parameter \cite{grangier} of $0.13\pm0.01$ without any background/dark-count subtraction. This implies that in our test we can genuinely associate the outcomes of the measurements to properties of single particles, avoiding classical wave-like analogies~\cite{Emary1}. The state of the photon is prepared (pre-selected) in the polarisation state $|\psi_A \rangle = \cos\alpha |H\rangle + \sin\alpha |V \rangle$ by means of a calcite polarising beam splitter (PBS) and a half-wave plate (HWP).

The use of a single-mode fibre (SMF) then prepares the transverse profile $\mathcal{F}(x,y)$ in a Gaussian shape of width $\sigma$, which ensures that the two directions can be used as distinct pointers for the weak measurements~\cite{Piacentini}. These operations are implemented by coupling the polarisation to the transverse position by means of the unitary transformations $\widehat{U}_x= \exp (-i g_x \widehat{I}_B \otimes \widehat{P}_x)$ and $\widehat{U}_y= \exp (-i g_y \widehat{I}_C \otimes \widehat{P}_y)$, where $\widehat{I}_C=|H \rangle\langle H|- |V \rangle\langle V|$ and $\widehat{I_B}=|\psi_\gamma \rangle\langle \psi_\gamma|- |\psi_\gamma^\bot \rangle\langle \psi_\gamma^\bot|$ is associated to an arbitrary direction for the linear polarisation: $|\psi_\gamma\rangle=\cos\gamma|H\rangle+\sin\gamma|V\rangle$ and $|\psi_\gamma^\bot\rangle=\sin\gamma|H\rangle-\cos\gamma|V\rangle$.
The operators $\widehat{P}_x$ and $\widehat{P}_y$ are the momenta associated to the $x$ and $y$ positions, respectively.
The interaction $\widehat{U}_x$ ($\widehat{U}_y$) is brought about by a 2-mm-long birefringent crystal whose extraordinary ($e$) optical axis lies in the $x$-$z$ ($y$-$z$) plane, at a $\pi/4$-angle with respect to the $z$ direction. Due to the spatial walk-off effect experienced by the photons, the two polarization paths get slightly separated along the $x$ ($y$) direction.
The actual interaction along the $I_B$ basis can be tuned by means of a HWP.
The condition $g_x^2/\sigma^2\simeq g_y^2/\sigma^2\ll1$ ensures that the back-action on the incoming state is negligible, {i.e.} the measurement operates in the weak regime~\cite{Piacentini}.
Along with the spatial walk-off, each birefringent crystal also induces a temporal walk-off and a possible polarization change, both to be compensated to avoid unwanted additional decoherence effects. We were able to do this by adding after each crystal a second birefringent crystal of properly chosen length (1.1 mm) with the optical axis along the $y$ ($x$) directions respectively, each mounted on a piezo-controlled rotator with 100 $\mu$rad nominal resolution, allowing to cancel the temporal walk-off avoiding unwanted circular components in the polarisation state due to the previous interaction.

\begin{figure}[h!]
\begin{center}
\includegraphics[width=\columnwidth]{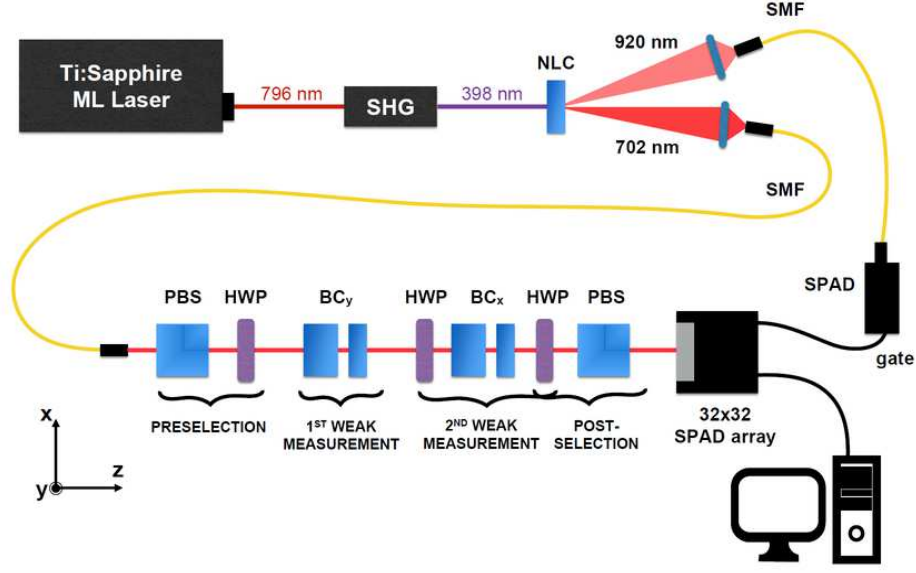}
\caption{Experimental setup. Heralded single photons are produced by downconversion in a 5-mm long LiIO$_3$ non-linear crystal (NLC) cut for type-I phasematching; the pump beam, obtained by second harmonic generation (SHG) of a modelocked laser (rep. rate 76MHz), produces idler ($\lambda_i=920$ nm) and signal ($\lambda_s=702$ nm) photons, which are then coupled in a single-mode fibre (SMF). The idler photons are detected by means of a Single-Photon Avalanche Diode (SPAD), which imparts a trigger to the signal detection system. Signal photons are prepared in the polarisation state $| \psi_{A}\rangle$ by means of a polarising beam splitter (PBS) and a half-wave plate (HWP) ($I_A=1$), then they pass through a birefringent system $BC_y$ that shifts them in the transverse $y$ direction, depending on their polarisation, thus measuring $I_C$ weakly. $BC_y$ consists of two birefringent crystals: the first one realizes the weak interaction, while the second one compensates temporal walk-off and decoherence effects. A similar system, $BC_x$, performs in cascade a weak measurement of $I_B$ by shifting the photons in the $x$ direction; this is placed after a HWP allowing to measure along an arbitrary linear polarization axis.  A further HWP is used to counter the basis change and decide the observable $I_D$ which determines the post-selection. The photons are finally detected by means of a spatial-resolving $32{\times}32$ SPAD array.
}
\label{setup}
\end{center}
\end{figure}

After the second weak interaction, the photons arrive to a HWP that undoes the preceding rotation and, at the same time, determines the projection of the state onto one of the post-selected states $\langle\psi_{A}|$, $\langle\psi_{D}| = \cos\delta \langle H| + \sin\delta \langle V |$ or $\langle \psi_{D}^\bot| = \sin\delta \langle H | - \cos\delta \langle V |$, by means of a PBS.

At the end of the optical path, the single photon is detected by a spatial-resolving single-photon detector prototype, i.e. a two-dimensional array made of $32\times32$ ``smart pixels''~\cite{villa} - each embedding a SPAD detector and its front-end electronics for counting and timing single photons - operating in parallel with a global shutter readout.
The SPAD array is operated in gated mode, with each count by the SPAD on the heralding arm triggering a 6 ns detection window in each pixel of the array. At our heralding rate of ${\sim}130$ kHz, the dark count rate of the array is drastically reduced by the low duty cycle, improving the signal-to-noise ratio.

Since we are interested in the LGT as a tool for probing quantumness, we estimate each term in the inequality \eqref{limit4} separately in our setup.
The chain of weak interactions and the space-resolved detector allow us to reconstruct the expectation values $\langle I_B \rangle$ and $\langle I_C \rangle$ by measuring the average $x$ and $y$ positions of the photons, respectively, when post-selecting on the input state $\langle \psi_A|$: $\langle \widehat{x} \rangle \simeq g_x \langle \widehat{I}_B \rangle $ and $\langle \widehat{y} \rangle \simeq g_y \langle \widehat{I}_C \rangle $.
The covariance of the $x$ and $y$ positions gives $\langle \widehat{x}\, \widehat{y} \rangle \simeq \frac{g_x g_y }{2} (\langle \widehat{I}_B \widehat{I}_C \rangle + \langle \widehat{I}_{B}  \rangle \langle \widehat{I}_C \rangle)$.
By inverting these relations, it is possible to obtain the single and sequential values $\langle \widehat{I}_C \rangle$, $\langle \widehat{I}_{B} \rangle$ and $\langle \widehat{I}_{B} \widehat{I}_C \rangle$, estimated as weak averages. This resolves a major difficulty, in that by using standard ``strong'' measurements one would only have access to the symmetrized  quantity $\frac{1}{2}\langle \psi_A|I_B I_C + I_C I_B|\psi_A\rangle$~\cite{sequential}.
Post-selection on $\langle\psi_{D}|$ and $\langle \psi_{D}^\bot|$ occurrence delivers the probabilities $p_D(1) = |\langle\psi_D|\psi_A\rangle|^2 $ and $p_D(-1) = |\langle\psi_D^\bot|\psi_A\rangle|^2$, as well as the weak values ${_{1}\ev{I_C}}$ and ${_{-1}\ev{I_C}}$.\\
\begin{figure}[h!]
\begin{center}
\includegraphics[width=0.95\columnwidth]{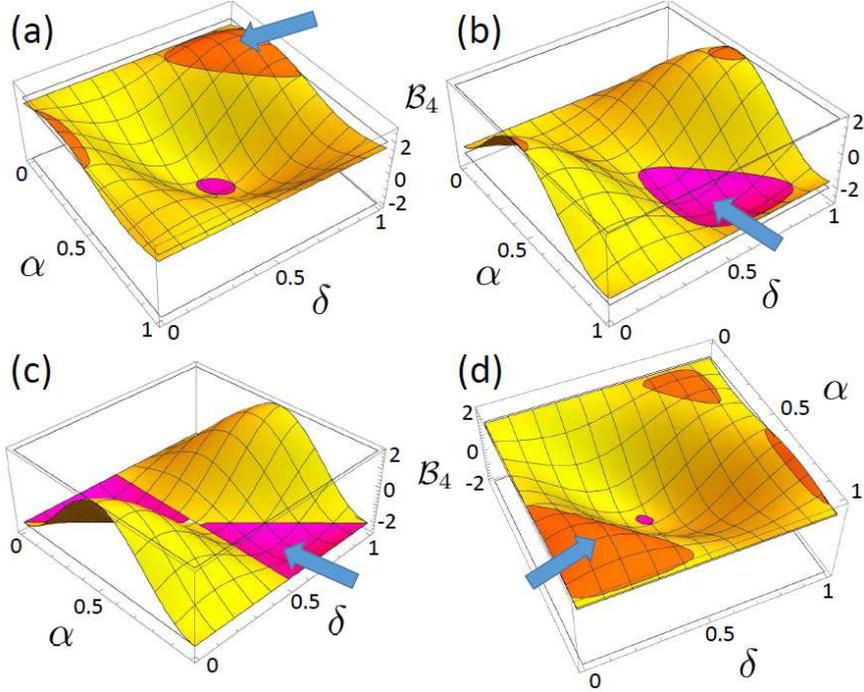}
\caption{Behaviour of the quantity $\mathcal{B}_{4}$ in eq. (\ref{limit4}) vs. the parameters $\alpha$ (related to the state $|\psi_A\rangle$) and $\delta$ (determining the states $|\psi_{D}\rangle$ and $|\psi_{D}^\bot\rangle$), both in $\pi$ units, for four different values of the parameter $\gamma$ defining the polarisation operator $I_B$: $\gamma=0.1\pi$ for plot (a), $\gamma=0.4\pi$ for plot (b), $\gamma=0.5\pi$ for plot (c), $\gamma=0.95\pi$ for plot (d). In each of these plots, the yellow part of the surface indicates the non-violation area ($-2 \leq \mathcal{B}_{4} \leq 2$), while in orange and magenta are highlighted respectively the positive ($\mathcal{B}_{4}>2$) and negative ($\mathcal{B}_{4}<-2$) violation areas. In each plot, the blue arrow indicates the point for which the violation was experimentally checked
}
\label{simul}
\end{center}
\end{figure}

{\it Results and conclusions.} Fig. \ref{simul} reports a theoretical simulation showing the shape of $\mathcal{B}_{4}$ for four different values of $\gamma$, plotted vs. the parameters $\alpha$ and $\delta$ determining the pre- and post-selection states.
Aside of the yellow part of the surface, indicating where the classical bound holds, for each $\gamma$ value one observes orange and/or magenta areas, corresponding to the $\mathcal{B}_{4} > 2$ and $\mathcal{B}_{4} < -2$ violations respectively.\\
We tested the inequality for different choices of the initial state $\alpha$, of the orientation $\gamma$ of weak measurement, and of the final post-selection $\delta$: the four combinations have been identified to deliver a violation (indicated by the blue arrow in each plot reported in Fig. \ref{simul}) close to the maximal value, whose results are illustrated in Table \ref{results}. For each of the four tests performed, the experimental values of $\mathcal{B}_{4}$ are in excellent agreement with the theoretical expectations within the statistical uncertainties, granting for both the positive and negative values a classical bound violation between 3.4 and 4.4 standard deviations. In the table, we also report the measured weak values showing how anomalies, {\it i.e.} values outside the standard range $-1$ to $1$, do flag the violation of the Leggett-Garg inequality: this corroborates the intimate connection between the emergence of anomalous values and the failure of a realistic description.

\begin{table}[htbp]
\begin{center}
\begin{tabular}{|c|c|c|c|c|}
\hline
\hspace{0.3mm} Parameters \hspace{0.3mm} & \hspace{0.1mm} $\mathcal{B}_{4}^{(th)}$ \hspace{0.1mm} & \hspace{1.5mm} $\mathcal{B}_{4}^{(exp)}$ \hspace{1.5mm} & \hspace{1mm} ${_1\langle I_c\rangle}$ \hspace{1mm} & ${_{-1}\langle I_c\rangle}$\\
\hline
$\gamma=0.1\pi$ & & & &\\
$\alpha=0.233\pi$ & $2.82$ & $2.76\pm0.17$ & $2.34\pm0.04$  & $-0.34\pm0.04$\\
$\delta=0.867\pi$ & & & &\\
\hline
$\gamma=0.4\pi$ & & & &\\
$\alpha=0.767\pi$ & $-2.82$ & $-2.74\pm0.18$ & $-0.30\pm0.04$ & $2.20\pm0.04$ \\
$\delta=0.633\pi$ & & & &\\
\hline
$\gamma=0.5\pi$ & & & &\\
$\alpha=0.833\pi$ & $-2.50$ & $-2.56\pm0.16$ & $0.01\pm0.06$ & $1.86\pm0.06$\\
$\delta=0.667\pi$ & & & &\\
\hline
$\gamma=0.95\pi$ & & & &\\
$\alpha=0.8\pi$ & 2.71 & $2.86\pm0.19$ & $1.86\pm0.04$ & $-0.12\pm0.06$ \\
$\delta=0.15\pi$ & & & &\\
\hline
\end{tabular}
\end{center}
\caption{Leggett-Garg inequality violation results obtained in our four experimental scenarios. The first column reports the $\gamma$, $\alpha$ and $\delta$ values exploited in each experiment, the second and third columns host respectively the theoretical ($\mathcal{B}_{4}^{(th)}$) and experimentally-obtained ($\mathcal{B}_{4}^{(exp)}$) values of the quantity $\mathcal{B}_{4}$, while the fourth and fifth columns show the anomalous weak values obtained for $I_C$ in each experiment.
}
\label{results}
\end{table}

We demonstrated the capability of our setup to address single photons with negligible disturbance, certifying it by a LGT and, in a complementary way, by the presence of anomalous weak values upon post-selection. This is a manifestation of the good quality of our device, which may find applications to random number generators~\cite{Curchod,Schiavon,Hu}.\\
\\
\\
This work has been supported by EMPIR-14IND05 ``MIQC2'' (the EMPIR initiative is co-funded by the EU H2020 and the EMPIR Participating States) and the MIUR Progetto Premiale 2014 ``Q-SecGroundSpace''.

\section{Appendix}
We can generalise the connection between the violation of a LG inequality and post-selected values as follows: consider a sequence of $m$ binary measurements $I_n$, with the first one $I_1$ coinciding with state preparation, hence $I_1=1$ deterministically. A generalised LG inequality is written as:
\begin{equation}
\label{piu}
-n\delta_{n=2k+1}-(n-2)\delta_{n=2k}\leq \mathcal{B}_n \leq n-2,
\end{equation}
being:
\begin{equation}
\label{piu1}
\mathcal{B}_n=\sum_{m=1}^{n-2}\langle I_m I_{m+1}\rangle+\langle I_{n-1} I_n\rangle - \langle I_{1} I_n\rangle.
\end{equation}
We now consider the last measurement $I_n$, and distinguish between the events for which $I_n=1$ or $I_n=-1$, each occurring with probabilities $p_+$ and $p_-$, respectively.
This leads us to consider the post-selected values ${_\pm\langle I_n I_{n+1}\rangle}$ for any correlator in \eqref{piu}:
\begin{equation}
\begin{aligned}
\label{piuPS}
\mathcal{B}_n=p_+\left( \sum_{m=1}^{n-2}{_+\langle I_m I_{m+1}\rangle}+{_+\langle I_{n-1} \rangle} -1 \right)\\
+p_-\left( \sum_{m=1}^{n-2}{_-\langle I_m I_{m+1}\rangle}-{_-\langle I_{n-1} \rangle} +1 \right)
\end{aligned}
\end{equation}
If we now assume that all post-selected values are regular, in that they are both within the spectrum of ordinary eigenvalues, the term in $p_+$ is upper bounded by $n-2$. The term in $p_-$ actually contains an expression akin to $\mathcal{B}_{n-1}$ for the post-selected values; the upper bound for the whole quantity is $n-2$, as well. Therefore, the regularity of the post-selected values in both their domain and their compatibility with macroscopic realism, leads to the LG inequality \eqref{piu}.


\begin{thebibliography}{99}
\bibitem{mg1} M.Genovese, {\it Adv. Sci. Lett.} {\bf 3}, 249 (2010)
\bibitem{Bell64} J.S. Bell, {\it Speakable ad Unspeakable in Quantum Mechanics} (Cambridge University Press, Cambridge (UK) 1987).
\bibitem{mg} M. Genovese, {\it Phys. Rep.} {\bf 413}, 319 (2005), and references therein.
\bibitem{KS39} S. Kochen, and E.P. Specker, {\it J. Math. Mech.} {\bf 17}, 59 (1967).
\bibitem{LG85} A.J. Leggett , and A. Garg, {\it Phys. Rev. Lett.} {\bf 54}, 857 (1985).

\bibitem{bruckner} J. Kofler and \v C. Brukner, {\it Phys. Rev. Lett.} {\bf 99}, 180403 (2007).
\bibitem{bruckner2} J. Kofler and \v C. Brukner, {\it Phys. Rev. Lett.} {\bf 10}1, 090403 (2008).
\bibitem{mm} M. Barbieri, {\it Phys. Rev. A} {\bf 80}, 034102 (2009).
\bibitem{Lambert} N. Lambert, C. Emary, Y.-N. Chen, and F. Nori, {\it Phys. Rev. Lett.} {\bf 105}, 176801 (2010).
\bibitem{Nimmrichter}  S. Nimmrichter, and K. Hornberger, {\it Phys. Rev. Lett.} {\bf 110}, 160403 (2013).
\bibitem{BuEmary} C. Budroni, and C. Emary, {\it Phys. Rev. Lett.} {\bf 113}, 050401 (2014).
\bibitem{Robens} C. Robens, W. Alt, D. Meschede, C. Emary, and A. Alberti., {\it Phys. Rev. X } {\bf 5}, 011003 (2015).
\bibitem{Morgan} C. Budroni, G. Vitagliano, G. Colangelo, R. J. Sewell, O. G\"uhne, G. T\'oth, and M. W. Mitchell, {\it Phys. Rev. Lett. } {\bf  115}, 200403 (2015).

\bibitem{Palacios} A. Palacios-Laloy, F. Mallet, F. Nguyen, P. Bertet, D. Vion, D. Esteve, and A. N. Korotkov,  {\it Nat. Phys.} {\bf  6}, 442 (2010).
\bibitem{Goggin} M. E. Goggin, M. P. Almeida, M. Barbieri, B. P. Lanyon, J. L. O'Brien, A. G. White, and G. J. Pryde, {\it Proc. Natl. Acad. Sci. USA } {\bf 108}, 1256 (2011).
\bibitem{Dressel} J. Dressel, C. J. Broadbent, J. C. Howell, and A. N. Jordan, {\it Phys. Rev. Lett.} {\bf  106}, 040402 (2011).
\bibitem{Athalye} V. Athalye, S. Singha Roy, and T. S. Mahesh, {\it Phys. Rev. Lett.} {\bf  107}, 130402 (2011)
\bibitem{Knee}  G. C. Knee, S. Simmons, E. M. Gauger, J. J. Morton, H. Riemann, N. V. Abrosimov, P. Becker, H.-J. Pohl, K. M. Itoh, M. L. Thewalt, G. A. D. Briggs, and S. C. Benjamin, {\it Nat. Commun.} {\bf  3}, 606 (2012).
\bibitem{Holger} Y. Suzuki, M. Iinuma, and H.F. Hofmann, {\it New J. Phys.} {\bf  14}, 103022 (2012)
\bibitem{Huelga} Z.-Q. Zhou, S.F. Huelga, C.-F. Li, and G.-C. Guo, {\it Phys. Rev. Lett.} {\bf  115}, 113002 (2015).
\bibitem{Knee2} G. C. Knee, K. Kakuyanagi, M.-C. Yeh, Y. Matsuzaki, H. Toida, H. Yamaguchi, S. Saito, A.J. Leggett, and W.J. Munro, {\it Nat. Comms } {\bf 7}, 13253 (2016).

\bibitem{Emary14} C. Emary, N. Lambert, and F. Nori, {\it Rep. Prog. Phys.} {\bf  77}, 016001 (2014) offers a comprehensive review of recent results.

\bibitem{Vedral} \v C. Brukner, S. Taylor, S. Cheung, and V. Vedral, arXiv quant-ph/042127 (2004).
\bibitem{Fritz} T. Fritz, {\it New J. Phys.} {\bf  12}, 083055 (2010).
\bibitem{Fedrizzi} A. Fedrizzi, M.P. Almeida, M.A. Broome, A.G. White, and M. Barbieri, {\it Phys. Rev. Lett. } {\bf 106}, 200402 (2011).
\bibitem{Budroni} C. Budroni, T. Moroder, M. Kleinmann, and O. G\"uhne, {\it Phys. Rev. Lett.} {\bf  111}, 020403 (2013).
\bibitem{Brierley} S. Brierley, A. Kosowski, M. Markiewicz, T. Paterek, and A. Przysiezna, {\it Phys. Rev. Lett.} {\bf  115}, 120404 (2015)

\bibitem{china} J.-S. Tang et al, {\it Chin. Phys. Lett.} {\bf  28}, 060304 (2011).
\bibitem{weak1} Y. Aharonov and L. Vaidman, {\it Phys. Rev. A } {\bf 41}, 11 (1990).
\bibitem{Wiseman} H.M. Wiseman,{\it  Phys. Rev. A } {\bf 65}, 032111 (2002).
\bibitem{Jordan} A.N. Jordan, A.N. Korotkov, and M. Buttiker, {\it Phys. Rev. Lett.} {\bf  97}, 026805 (2006)
\bibitem{Jordan1} N.S. Williams, and A.N. Jordan, {\it Phys. Rev. Lett.} {\bf 100}, 026804 (2008).

\bibitem{Ferrie} C. Ferrie, and J. Combes, {\it Phys. Rev. Lett.} {\bf  113}, 120404  (2014).
\bibitem{Brodutch} A. Brodutch, {\it Phys. Rev. Lett} {\bf . 114}, 118901 (2015).
\bibitem{FerrieR} C. Ferrie, and J. Combes, {\it Phys. Rev. Lett.} {\bf  114}, 118902 (2015).
\bibitem{Aharonov} Y. Aharonov, and D. Rohrlich, arXiv:1410.0381 (2014).



\bibitem{AharonovS} Y. Aharonov, D.Z. Albert, and L. Vaidman, {\it Phys. Rev. Lett.} {\bf  60}, 1351 (1988).
\bibitem{Pryde} G. J. Pryde, J. L. O'Brien, A. G. White, T. C. Ralph, and H. M. Wiseman, {\it Phys. Rev. Lett.} {\bf  94}, 220405 (2005).
\bibitem{JordanErr} N.S. Williams, and A.N. Jordan, {\it Phys. Rev. Lett.} {\bf  103}, 089902 (2009).

\bibitem{pusey} M. Pusey, {\it Phys. Rev. Lett.} {\bf  113}, 200401 (2014).
\bibitem{pusey2} F. Piacentini, A. Avella, M. P. Levi, R. Lussana, F. Villa, A. Tosi, F. Zappa, M. Gramegna,
G. Brida, I. P. Degiovanni, and M. Genovese, {\it Phys. Rev. Lett.} {\bf  116}, 180401 (2016).

\bibitem{Piacentini}  F. Piacentini, M. P. Levi, A. Avella, E. Cohen, R. Lussana, F. Villa, A. Tosi, F. Zappa, M. Gramegna, G. Brida, I. P. Degiovanni, and M. Genovese, {\it Phys. Rev. Lett.} {\bf  117}, 170402 (2016).
\bibitem{CHSH} J.F. Clauser, R.A. Holt,  A. Shimony, and M.A. Horne, {\it Phys. Rev. Lett.} {\bf  23}, 880 (1969).

\bibitem{IvoOE} S. Castelletto, I. P. Degiovanni, V. Schettini, and A. Migdall, {\it Opt. Expr.} \textbf{13} (18), 6709 (2005). 
\bibitem{grangier} P. Grangier, G. Roger, A. Aspect, {\it Eur. Phys. Lett.} \textbf{1}, 173 (1986). 
\bibitem{Emary1} J.-S. Xu, C.-F. Li, X.-B. Zou, and G.-C. Guo, {\it Sci. Rep.} {\bf 1}, 101 (2011).

\bibitem{villa} F. Villa, R. Lussana, D. Bronzi, S. Tisa, A. Tosi, F. Zappa, A. Dalla Mora, D. Contini, D. Durini, S. Weyers, and W. Brockherde,  {\it IEEE J. Sel. Top. Quantum Electron.} \textbf{20}, 3804810 (2014). 

\bibitem{sequential} G. Mitchison, R. Jozsa, and S. Popescu,  \emph{Phys. Rev. A} \textbf{76}, 062105 (2007). 


\bibitem{Curchod} F.J. Curchod, M. Johansson, R. Augusiak, M.J. Hoban, P. Wittek, andA. Ac\'in, {\it Phys. Rev. A} {\bf 95} 020102 (2012).
\bibitem{Schiavon} M. Schiavon, L. Calderaro, M. Pittaluga, G. Vallone, and P. Villoresi, {\it Quantum Sci. Technol.} {\bf 2}, 015010 (2017).
\bibitem{Hu} M.J. Hu, et al., arXiv:1609.01863 (2016).





\end{thebibliography}
\end{document}